\documentclass{aa}
\usepackage{times,float}
\usepackage{epsfig}
\usepackage{graphics}
\usepackage{amssymb}

\begin{document}

\title{A   possible   bright   blue    SN   in   the   afterglow   of   GRB
020305\thanks{Based on observations made with the Nordic Optical Telescope,
operated on  the island of La  Palma jointly by  Denmark, Finland, Iceland,
Norway, and Sweden.}}

\titlerunning{The host galaxy and the blue afterglow of GRB~020305}

\author{
    J. Gorosabel \inst{1,2}
    \and J.P.U. Fynbo \inst{3}
    \and A. Fruchter \inst{2}
    \and A. Levan \inst{4}
    \and J. Hjorth \inst{3}
    \and P. Nugent \inst{5}
    \and A.J. Castro-Tirado \inst{1}
    \and J.M. Castro Cer\'on \inst{2,3}
    \and J. Rhoads \inst{2}
    \and D. Bersier \inst{2}
    \and I. Burud \inst{2}
}

\offprints{J. Gorosabel}

\institute{
     Instituto de Astrof\'{\i}sica de Andaluc\'{\i}a (IAA-CSIC),
     P.O. Box 03004, E-18080 Granada, Spain
     \and
     Space Telescope Science Institute, 3700 San Martin Drive,
     Baltimore, MD 21218, USA
     \and 
     Niels   Bohr  Institute,   University  of
     Copenhagen, Juliane Maries Vej 30, DK-2100 K\o benhavn \O, Denmark
     \and     
     Department of Physics and Astronomy, University of Leicester,
     University Road, Leicester, LE1 7RH, UK
     \and     
     Lawrence Berkeley National Laboratory, MS 50-F, 1 Cyclotron Road,
     Berkeley, CA 94720, USA
     }
\mail{jgu@iaa.es}

\date{Received ; accepted }

\abstract{We report on ground-based and HST(+STIS) imaging of the afterglow
and host  galaxy of  the Gamma-Ray Burst  (GRB) of  March 5 2002.   The GRB
occurred  in a  $R=25.17\pm0.14$ galaxy,  which  apparently is  part of  an
interacting  system.   The lightcurve  of  the  optical  afterglow shows  a
rebrightening,  or at  least a  plateau,  12--16 days  after the  gamma-ray
event.  $UBVRIK^{\prime}$ multi-band imaging of the afterglow $\sim$12 days
after the GRB  reveals a blue spectral energy  distribution (SED).  The SED
is   consistent    with   a   power-law   with   a    spectral   index   of
$\beta=-0.63\pm0.16$, but  there is tentative evidence  for deviations away
from  a power-law.  Unfortunately,  a spectroscopic  redshift has  not been
secured for GRB~020305.   From the SED we impose a  redshift upper limit of
$z\lesssim2.8$, hence  excluding the pseudo redshift of  $4.6$ reported for
this burst.   We discuss the  possibilities for explaining  the lightcurve,
SED  and   host  galaxy  properties  for  GRB~020305.    The  most  natural
interpretation of  the lightcurve  and the SED  is an  associated supernova
(SN).  Our data  can not precisely determine the redshift  of the GRB.  The
most favoured explanation  is a low redshift ($z\sim0.2$)  SN, but a higher
redshift ($z  \gtrsim 0.5$) SN  can not be  excluded. We also  discuss less
likely scenarios  not based  on SNe,  like a burst  occurring in  a $z=2.5$
galaxy  with  an  extinction  curve  similar  to that  of  the  Milky  Way.
 }

\maketitle

\section{Introduction}

For long  duration GRBs  the relation with  supernovae (SNe)  became firmly
established  with  the  discovery   of  the  type  Ic  supernova  SN~2003dh
associated  with GRB~030329 (Stanek  et al.   \cite{Stan03}; Hjorth  et al.
\cite{Hjor03}).  This  result lends strong  support to the  collapsar model
(Woosley \cite{Woos93}),  but a  SN is also  an ingredient in  other models
(e.g.   Dado  et  al.    \cite{Dado03};  Fryer  \&  Heger  \cite{Frye04}).
However,  the  associated  SNe  follow  a  broad  distribution  of  optical
luminosities  (Zeh et  al.  \cite{Zeh04}).   Furthermore the  connection of
GRBs with SNe of other types than  Ic can not be excluded, motivated by the
two possible  associations of GRBs and  II type SNe  (SN~1997cy, Germany et
al.  \cite{Germ00}; SN~1999E, Rigon et al.  \cite{Rigo03}).  Therefore, the
afterglow  lightcurves and  SEDs  around the  SN  peak are  far from  being
described by an universal SN template.  In this study we present ground and
space-based  optical observations  of GRB~020305 carried  out from  11.5 to
321.2 days after the burst.

GRB~020305  was localised  by  the  HETE-II satellite  on  March 5.4968  UT
(Ricker et  al.  \cite{Rick02}).  The  high-energy emission as seen  by the
Interplanetary network  (IPN) consisted in  two broad pulses, with  a total
GRB duration of  $\sim$280s (Hurley et al.   \cite{Hurl02}), placing it in
the long-soft  burst category.  Price et al.   (\cite{Pric02}) reported the
presence  of a transient  optical source  in the  HETE-II/IPN error  box in
images taken $\sim$20  hours after the GRB.  Further  imaging confirmed the
fading  behaviour of  the candidate  (Lee et  al.  \cite{Lee02};  Ohyama et
al. \cite{Ohya02}).

The paper  is structured as follows: Sect.   \ref{observations} details the
observations  and  the data  reduction,  Sect.   \ref{results} reports  the
results  on the SED,  lightcurve and  host galaxy,  Sect.  \ref{discussion}
discusses  several  interpretations  of  the  results,  and  finally  Sect.
\ref{conclusions} draws the conclusions of this study.

\section{Observations and data reduction}
\label{observations}
\subsection{NOT observations}
We observed the field of GRB~020305  from the ground with the 2.56-m Nordic
Optical Telescope  (NOT) on 2002 March 16.95--22.18  UT, i.e.  11.45--16.68
days  after the  GRB.  The  instrument  used was  the Andaluc\'{\i}a  Faint
Object  Spectrograph (ALFOSC)  equipped  with a  $2048^2$  pixel Loral  CCD
having a pixel  scale of $0\farcs189$.  The data  were reduced using standard
methods. The $UBVRI$-band calibration was carried out observing the Landolt
field  PG1047+003 (Landolt  \cite{Land92}) with  the 3.58-m  New Technology
Telescope (NTT) on 2003 February 28.

\begin{figure}[t]
\begin{center}
  {\includegraphics[width=\hsize]{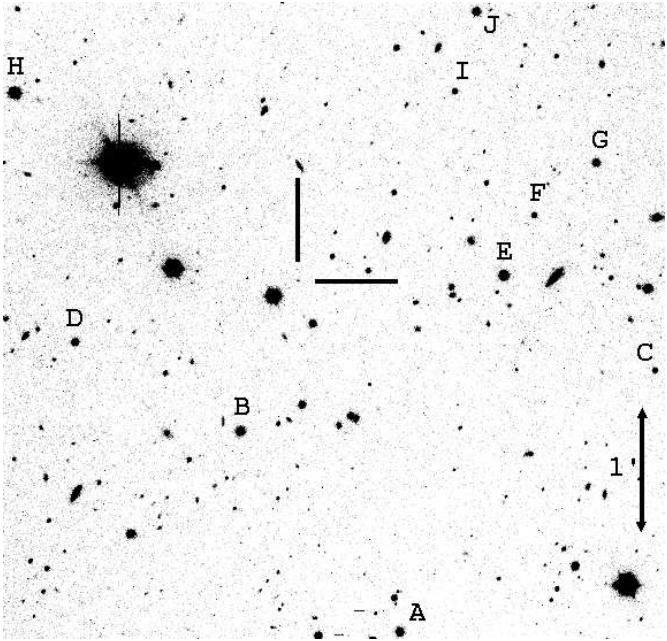}}
\caption{\label{standares} The co-added NOT  R-band image taken on 2002 Mar
  16.993-17.039 UT.  The tickmarks indicate the position of the OA, and the
  labels  the  standard stars  listed  in Table~\ref{tablestandares}.   The
  scale of  the image is indicated with  a vertical bar at  the lower right
  corner. North is upwards and East is leftwards.}
\end{center}
\end{figure}

Fig.~\ref{standares} shows an image  of the optical afterglow (OA) and
the   positions   of   the   secondary  standard   stars   listed   in
Table~\ref{tablestandares}.   The mean celestial  coordinates obtained
from  6 NOT  images are;  R.A.(J2000)= $12^{h}  42^{m}  27.963^{s} \pm
0.020^{s}$,  Dec(J2000)=  $-14^{\circ}18^{\prime}11.45^{\prime \prime}
\pm 0\farcs20$.  The astrometry is  based on $\sim30$ USNO~A2.0 stars
per image.   The errors do  not include the systematic  uncertainty of
the    USNO~A2.0    catalogue    ($\sim0\farcs25$;    Assafin    et
al. \cite{Assa01}).

\begin{table*}
\begin{center}
\caption{Calibrated  magnitudes   of  the  secondary   standard  stars
 displayed in  Fig.~\ref{standares}. For the faintest  standards (C, F
 and I) the  $U$-band magnitudes show large errors  ($>0.08$ mag) so the
 corresponding field is empty. }
\begin{tabular}{lccccccc}
\hline
 ID& R.A.(J2000)  & Dec(J2000)                 &$U$&$B$&$V$&$R$&$I$\\
  & $h~~m~~s$   & $\circ~~\prime~~\prime\prime$&   &   &   &   &   \\
\hline
A &$12:42:24.45$&$-14:21:02.7$&$18.56\pm0.03$&$18.48\pm0.05$&$17.72\pm0.03$&$17.27\pm0.04$&$16.82\pm0.03$\\
B &$12:42:29.83$&$-14:19:25.6$&$18.88\pm0.04$&$18.45\pm0.05$&$17.58\pm0.03$&$17.08\pm0.03$&$16.68\pm0.03$\\
C &$12:42:15.98$&$-14:18:53.0$&$ ------     $&$20.62\pm0.05$&$19.91\pm0.03$&$19.46\pm0.03$&$19.06\pm0.04$\\
D &$12:42:35.41$&$-14:18:43.3$&$18.58\pm0.04$&$18.80\pm0.04$&$18.32\pm0.03$&$17.93\pm0.03$&$17.59\pm0.03$\\
E &$12:42:21.01$&$-14:18:08.2$&$17.54\pm0.04$&$17.54\pm0.04$&$16.94\pm0.02$&$16.58\pm0.03$&$16.21\pm0.03$\\
F &$12:42:20.03$&$-14:17:38.7$&$ ------     $&$20.83\pm0.06$&$19.70\pm0.03$&$19.00\pm0.04$&$18.36\pm0.03$\\
G &$12:42:17.95$&$-14:17:12.6$&$20.64\pm0.07$&$19.47\pm0.05$&$18.41\pm0.03$&$17.79\pm0.04$&$17.27\pm0.03$\\
H &$12:42:37.55$&$-14:16:40.9$&$17.73\pm0.03$&$17.40\pm0.04$&$16.72\pm0.02$&$16.24\pm0.03$&$15.88\pm0.03$\\
I &$12:42:22.80$&$-14:16:38.0$&$ ------     $&$19.61\pm0.04$&$19.17\pm0.03$&$18.87\pm0.04$&$18.42\pm0.03$\\
J &$12:42:22.16$&$-14:15:59.1$&$18.14\pm0.03$&$18.28\pm0.04$&$17.74\pm0.03$&$17.39\pm0.03$&$17.04\pm0.03$\\

\hline
\label{tablestandares}
\end{tabular}
\end{center}
\end{table*}

\begin{table}
\begin{center}
\caption{{\it    Upper   subtable:}   Log    of   the    ground-based   NOT
  observations. The  magnitudes are  given in the  Vega system and  are not
  corrected for Galactic extinction.  The contribution of the host has
  not been subtracted.  {\it Lower subtable:} Log of the HST observations.
  The HST magnitudes  are in the AB system, are  not corrected for Galactic
  extinction and only consider the OA flux.}
\begin{tabular}{lcccc}
\hline
Date    & Filter & Exp.Time & Mag. & \hspace{-0.3cm}Seeing\\
2002 UT   &    &   (s) &   & \hspace{-0.3cm}$\prime\prime$\\
\hline
{NOT(+ALFOSC)}\\
Mar $16.950$-$16.993$&$I$&$9 \times300 $&$22.43\pm0.11$&$\hspace{-0.3cm}1.0$\\
Mar $16.993$-$17.039$&$R$&$4 \times900 $&$22.76\pm0.05$&$\hspace{-0.3cm}1.0$\\
Mar $17.039$-$17.084$&$V$&$3 \times1200$&$22.88\pm0.07$&$\hspace{-0.3cm}1.3$\\
Mar $17.098$-$17.173$&$U$&$5 \times1200$&$22.75\pm0.14$&$\hspace{-0.3cm}1.4$\\
Mar $17.181$-$17.226$&$B$&$3 \times1200$&$23.25\pm0.11$&$\hspace{-0.3cm}1.7$\\
Mar $18.960$-$19.040$&$R$&$7 \times900 $&$22.48\pm0.08$&$\hspace{-0.3cm}2.4$\\
Mar $20.993$-$21.164$&$R$&$15\times900 $&$22.42\pm0.06$&$\hspace{-0.3cm}2.2$\\
Mar $21.981$-$22.180$&$R$&$15\times900 $&$22.41\pm0.05$&$\hspace{-0.3cm}1.9$\\
\hline        
{HST(+STIS)}\\   
Apr $12.667$-$12.888$&CL& $7402^{\star}  $ & $24.80\pm0.02$ &\hspace{-0.3cm}-\\
Apr $14.135$-$14.412$&LP& $7567^{\star}  $ & $24.45\pm0.03$ &\hspace{-0.3cm}-\\
Jun $16.454$-$16.550$&CL& $4856^{\dagger}$ & $27.86\pm0.20$ &\hspace{-0.3cm}-\\
Jun $16.597$-$16.751$&LP& $5021^{\dagger}$ & $26.98\pm0.22$ &\hspace{-0.3cm}-\\
Dec $13.832$-$13.998$&LP& $7966^{\star}  $ & $>28.07$       &\hspace{-0.3cm}-\\
Jan $20.534$-$20.695^{\ddagger}$&CL& $7306^{\star}$&$>28.20$&\hspace{-0.3cm}-\\
\hline      
\multicolumn{4}{l}{$\star$ Composed by 9 exposures of different durations.}\\
\multicolumn{4}{l}{$\dagger$ Composed by 6 exposures of different durations.}\\
\multicolumn{4}{l}{$\ddagger$ 2003.}\\
\hline
\label{obs-journal}
\end{tabular}
\end{center}
\end{table}

\subsection{HST observations}

 From  space  the  OA  was  observed  using  the  Space  Telescope  Imaging
 Spectrograph  (STIS)  on  the  Hubble  Space  Telescope  (HST).   The  HST
 observations  were  carried  out   clustered  around  three  mean  epochs;
 $\sim39$,  $\sim103$ and $\sim300$  days after  the GRB.   In each  of the
 three epochs two broad filters were used; the long-pass (LP) and the 50CCD
 clear  (CL)   filter.   The  journal  of  observations   is  displayed  in
 Table~\ref{obs-journal}.

 The CCD gain was set to 1 $e^{-}$/ADU and the read-out noise was 4.46
 $e^{-}$.  The STIS images were preprocessed through the standard STIS
 pipeline and  combined using the DITHER (v1.2)  software (Fruchter \&
 Hook         \cite{Fruc02})         as         implemented         in
 IRAF~(v2.11.3)/STSDAS~(v2.1.1).  The `pixfrac'  parameter was  set to
 0.7  and  the  output  scale  to  $0\farcs0254$  pixel$^{-1}$  ('scale'
 parameter 0.5).

\begin{figure*}[t]
\begin{center}
{\includegraphics[width=8.9cm]{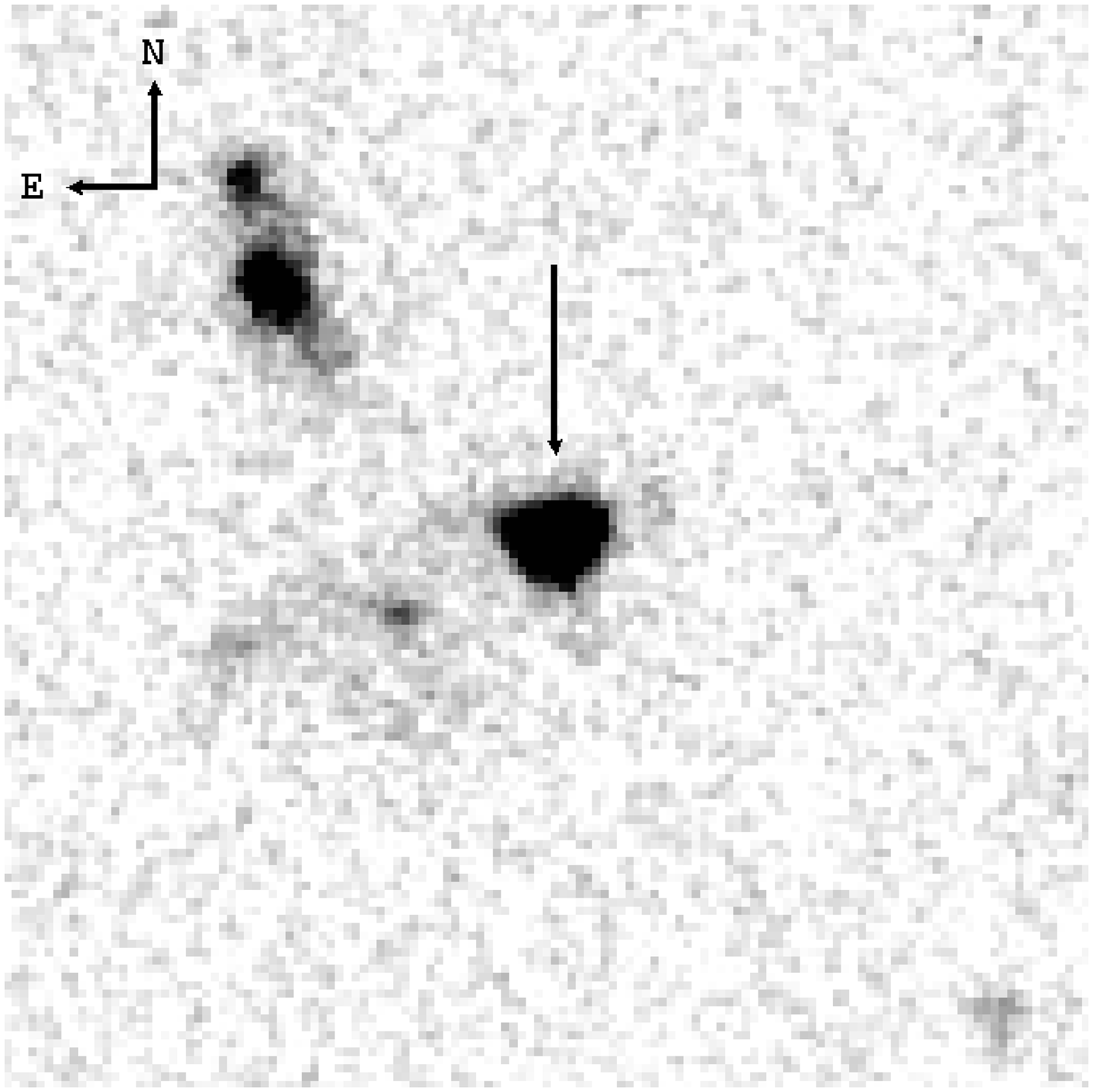}}
{\includegraphics[width=8.9cm]{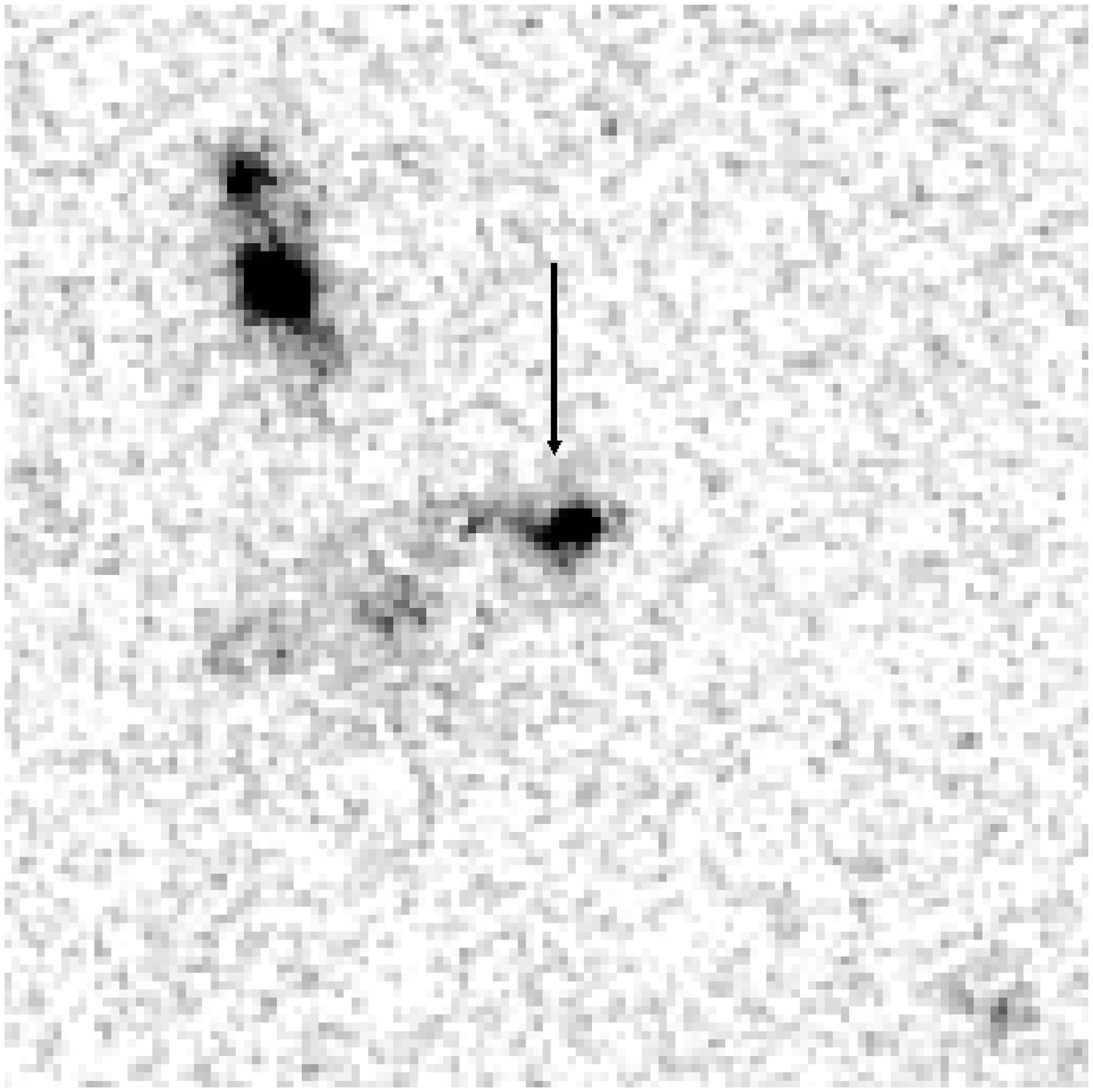}}
\caption{\label{hst}  Fading of  the GRB~020305  optical  afterglow as
  seen  in the CL  band.  The  observations were  carried out  on 2002
  April 12 (left panel) and 2003 January 20 (right panel).  The images
  are centred  on the  afterglow (indicated by  an arrow) and  cover a
  field of view  of $3\farcs3 \times 3\farcs3$.  We  note the presence
  of a bright knot $\sim  0\farcs07$ ($\sim 3$ pixels) westwards and a
  potential interacting galaxy $\sim 1\farcs1$ northeastwards from the
  OA position.  North is upwards and East is leftwards.}
\end{center}
\end{figure*}

 For  both STIS filters  the third  epoch images  were subtracted  from the
 first and second epoch frames, allowing us to perform photometry of the OA
 with  no contamination  from  the host  galaxy.   Aperture photometry  was
 carried  out with  a small  aperture radius  of 4  drizzled  pixels ($\sim
 0.1^{\prime \prime}$), which gave the optimal signal-to-noise ratio.  Then
 an aperture  correction was applied for  both filters based  on the growth
 curve  of   the  star  $CPD-60^{\circ}7585$\footnote{STIS   handbook  {\tt
 http://www.stsci.edu/hst/stis-   /documents/handbooks/}},   which  has   a
 similar $m^{AB}_{\rm  CL}-m^{AB}_{\rm LP}$ colour  to the OA.  The  use of
 other apertures yielded consistent aperture corrected magnitudes, implying
 that the assumed growth curve is suitable for the OA. Fig.~\ref{hst} shows
 the OA fading as imaged by HST.

\section{Results}
\label{results}
 
\subsection{The afterglow SED $\sim11.5$ days after the burst}
\label{sed}

The optical multicolour imaging carried  out during the first NOT observing
night (March 16.950--17.226 UT,  see Table~\ref{obs-journal}) allowed us to
construct  the optical SED  of the  afterglow.  Furthermore,  including the
$K^{\prime}$-band  detection of  March 14.39  UT reported  by Burud  et al.
(\cite{Buru02}) the SED  was extended to the near-infrared  (NIR).  We have
re-reduced and analysed the  original $K^{\prime}$-band data taken by Burud
et al.  (\cite{Buru02}).  In order  to enhance the signal-to-noise ratio we
convolved the final image with  the Point Spread Function (PSF) constructed
using field  stars. We obtained  a value of  $K^{\prime} = 20.10  \pm 0.31$
consistent with  the value of  Burud et al.  (\cite{Buru02};  $K^{\prime} =
19.8\pm0.2$). 

The  extended underlying flux distribution  imposed a seeing-dependent
host  galaxy  subtraction.   Thus,   for  the  images  acquired  in  seeing
conditions above $1\farcs1$, the flux of the host plus the companion galaxy
(see Fig.~\ref{hst}) was subtracted. We  note that the seeing values of the
observations   by   Burud   et   al.   (\cite{Buru02}),   Ohyama   et   al.
(\cite{Ohya02})  and  Lee  et  al.   (\cite{Lee02})  were;  $\sim1\farcs6$,
$\sim0\farcs9$  and  $\sim3\farcs0$,   respectively  (Priv.   Comm.).   The
determination of the $UBVIK^{\prime}$-band  magnitudes for the host and the
companion galaxy  was based  on the $R$-band  magnitudes and  the power-law
spectral indexes reported in Sect.~\ref{hostgalaxy}.

 The  mentioned optical/NIR  data where  extrapolated in  time to  a common
epoch of March 17.0 UT,  assuming a power-law lightcurve ($F_{\nu} \sim
t^{-\alpha}$).   Two different values  of the  decay index  ($\alpha$) were
used  for the  $K^{\prime}$  and for  the  optical bands.   Given that  the
$K^{\prime}$-band measurement was carried out before the rebrightening, the
$\alpha_{K^{\prime}}$  value  was  obtained  assuming a  power-law  stretch
connecting the  $R$-band point  by Ohyama et  al. (\cite{Ohya02})  with our
$R$-band    magnitude     of    March    16.993--17.039     UT,    yielding
$\alpha_{K{^\prime}}=0.36$.  For  the rest  of bands ($UBVRI$)  the adopted
value of  $\alpha_{UVBRI}$ was obtained based on  our $R$-band measurements
of  March   16.993--17.039  UT   and  March  18.960--19.040   UT,  implying
$\alpha_{UBVRI}=-1.05$.  Thus, the $K^{\prime}$-band flux extrapolation was
performed  assuming a slow  fading, whereas  the $UBRVRI$-band  fluxes were
shifted in  time adopting a  rising lightcurve.  In any  case, considering
that the  data points were taken close  from March 17.0 UT, the final
SED is basically insensitive to the assumed $\alpha$ value.

The  magnitudes  have been  transformed  to  flux  densities following  the
conversion  factors given  by Fukugita  et al.   (\cite{Fuku95})  and Allen
(\cite{Alle00})  for the optical  and NIR  bands respectively.   The fluxes
were  subsequently  corrected  for  foreground Galactic  extinction,  which
amounts  to $E(B-V)=0.053$  according to  Schlegel et  al. (\cite{Sche98}).
The derived SED  can be seen in Figs.~\ref{sed1}  and ~\ref{sed2}.  The SED
is  formally  consistent  with  a  power-law  ($F_{\nu}  \sim  \nu^{\beta},
\beta=-0.63\pm0.16,  \chi^2/{\rm  d.o.f} =  0.8$)  but  there is  tentative
evidence for  deviations away from  a power-law, most notably  an $RI$-band
decrement.

The mere  NOT $U$-band  detection imposes an  upper limit to  the redshift.
Applying  a Lyman forest  blanketing model  to the  pure power-law  SED fit
(Madau \cite{Mada95}), and  convolving it with the ALFOSC  $U$-band and CCD
sensitivity curves, we  derived an upper limit of  $z\lesssim2.8$.  This is
inconsistent with  the reported  pseudo redshift based  on the  high energy
properties of the GRB ($z=4.6$, Atteia \cite{Atte03}).

\subsection{The $R$-band lightcurve}
\label{lightcurve}

The combination  of the NOT  and HST observations  (see Sect.~\ref{hstobs})
presented  here   and  data   points  from  other   authors  (Lee   et  al.
\cite{Lee02}; Ohyama et al.  \cite{Ohya02})  allowed to us to construct the
$R$-band lightcurve of the OA.   The $R$-band magnitudes reported by Ohyama
et al.   (\cite{Ohya02}) have  been shifted to  our photometric  zero point
(Kosugi \cite{Kosu03}).  The data point by Price et al. (\cite{Pric02}) has
not  been included due  to lack  of information  on the  photometric system
(their  detection  is  based  on  unfiltered  images)  and  the  unreported
error-bar.

The  resulting  OA  lightcurve  (see  Fig.~\ref{lightcurve1})  can  not  be
described   using   only  a   power-law   ($F_{\nu}  \sim   t^{-\alpha}$;
$\alpha=1.02\pm0.04, \chi^2/{\rm d.o.f}  = 34.9$).  The poor fit  is due to
the fast decay  between the first two HST epochs  and especially because of
the lightcurve bump present 12 -- 16 days after the GRB.

\subsection{The late time decay}
\label{hstobs}

The first two HST epochs imply a fast decay. A power-law fit leads to decay
slopes of $\alpha_{\rm  CL}=2.85\pm0.19$ and $\alpha_{\rm LP}=2.45\pm0.22$.
Extrapolating  this  decay to  the  third  epoch,  afterglow magnitudes  of
$m^{AB}_{\rm  CL}=31.38\pm0.43$   and  $m^{AB}_{\rm  LP}=29.66\pm0.46$  (AB
system) are  predicted.  This is consistent  with the fact  that no obvious
point source emission is seen at the position of the afterglow in the third
epoch  images.  For  the  third epoch  images  the peak  flux  of the  mean
azimuthal host galaxy profiles at  the afterglow position are equivalent to
$m^{AB}_{\rm CL} = 28.20$ and  $m^{AB}_{\rm LP} = 28.07$.  These magnitudes
can be  considered as conservative upper  limits of the  OA contribution in
the third HST epoch and have been included in Table~\ref{obs-journal}.

The broad STIS CL and LP filters only allow a rough determination of the OA
colour. In order to estimate the colours of the two first HST observations,
the CL and LP magnitudes have been  rescaled to April 13.5 UT and June 16.5
UT using the above calculated decay indexes ($\alpha_{\rm CL}=2.85\pm0.19$,
$\alpha_{\rm LP}=2.45\pm0.22$).  In any case, the first four HST visits are
well clustered around the two mentioned epochs, so the magnitude shifts are
$<  0.06$   mag.   The   $m^{AB}_{\rm  CL}-m^{AB}_{\rm  LP}$   colours  are
$0.46\pm0.04$  and  $0.88\pm0.30$, on  April  13.5  UT  and June  16.5  UT,
respectively  (uncorrected for  Galactic extinction).   This  is consistent
($1.39 \sigma$) with an achromatic decay  between 39 and 103 days after the
GRB.

The HST magnitudes on April 13.5 UT  and June 16.5 UT can be transformed to
the Vega system  using the LP and CL filter  throughputs implemented in the
SYNPHOT package  of IRAF  yielding $R=24.38 \pm  0.05$ (April 13.5  UT) and
$R=28.31^{+2.41}_{-1.07}$ (June 16.5 UT).

\subsection{The host galaxy}
\label{hostgalaxy}

 The HST observations revealed an extended galaxy ($\sim1.3^{\prime \prime}
 \times  0.6^{\prime  \prime}$)  coincident   with  the  OA  position  (see
 Fig.~\ref{hst}).   The integrated CL  and LP  magnitudes (measured  in the
 third epoch  observations) of  the host are  $m^{AB}_{\rm CL} =  25.38 \pm
 0.05$  and $m^{AB}_{\rm LP}  = 25.18  \pm 0.12$,  corresponding to  a Vega
 $R$-band magnitude of $R=25.17\pm0.14$.  The photometry was performed with
 SExtractor using the automatic aperture magnitude ({\tt MAG\_AUTO}), which
 is  suitable to  measure the  total flux  of extended  objects  (Bertin \&
 Arnouts \cite{Bert96}).  The STIS host magnitudes correspond to a spectral
 index   $\beta   =   -0.45^{+0.52}_{-0.43}$   (dereddened   for   Galactic
 extinction).

 The  HST  imaging  also  revealed  another  extended  source  located
  $1\farcs1$ northeastwards  from the host  (see Fig.~\ref{hst}).  The
  magnitudes of this galaxy are $m^{AB}_{\rm CL} = 25.14 \pm 0.03$ and
  $m^{AB}_{\rm LP} = 24.92\pm0.06$, corresponding to $R= 24.95\pm0.08$
  in  the  Vega  system.    The  inferred  spectral  index,  $\beta  =
  -0.53^{+0.26}_{-0.24}$, is similar to that of the host galaxy.

\section{Discussion}
\label{discussion}

A  natural  interpretation of  the  lightcurve and  the  SED  would be  the
existence of a SN component present  $\sim2$ weeks after the GRB. Given the
limited  number of  measurements  in the  lightcurve/SED,  and the  unknown
redshift,  the space  of  SN-like lightcurve/SED  solutions (determined  by
$A_{V}$, the SN  amplitude, the stretching factor, and  the temporal offset
between the SN  and GRB) is highly degenerate. Hence,  we will only attempt
to  demonstrate existence and  not uniqueness  of possible  SN fits  to our
data. For  simplicity we decided to fix  the stretching factor to  1 and to
assume that the  GRB event was simultaneous to the  SN.  Moreover, the blue
colour of the OA suggests that  the host galaxy extinction can not be high,
so we considered no host galaxy extinction ($A_{V}=0$).

\begin{figure}[t]
\begin{center}
\resizebox{\hsize}{!}{\includegraphics[bb= 30 22 527 463]{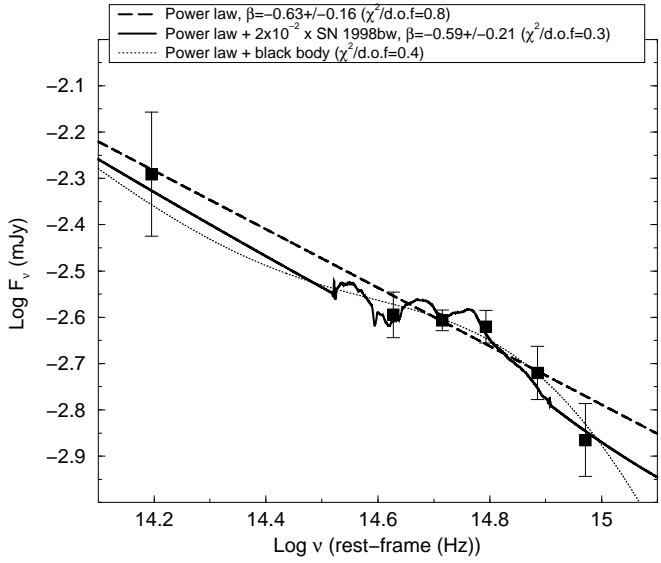}}
  \caption{\label{sed1} The  low redshift SN scenario.   The filled squares
  show the  $UBVRIK^{\prime}$-band data  points measured around  March 17.0
  UT.  The straight  long dashed line shows a pure  power-law fit ($\beta =
  -0.63 \pm  0.16$, $\chi^2/{\rm  d.o.f} = 0.8$),  whereas the  solid bumpy
  curve shows the fitted solution when a SN~1998bw-like component is added.
  The best SED solution is  obtained with a low redshift ($z\sim0.1$) faint
  SN (the amplitude is only $\sim2\%$ of SN~1998bw), which improves the fit
  ($\chi^2/{\rm  d.o.f}  =  0.3$).   Larger redshifts  make  the  SN~1998bw
  template unable to fit the high  $UB$-band flux, so are not plotted here.
  The dotted  thin smooth curve shows  the fit obtained  at $z\sim0.1$ when
  the contribution  of a  thermal spectrum and  a pure power-law  ($\beta =
  -1$) are added.  The rest-frame temperature  of the black body is $T \sim
  9200\pm1300$ K.  All  the fits assume no intrinsic  extinction, since the
  introduction of it provides worse fits.}
\end{center}
\end{figure}

\begin{figure}[t]
\begin{center}
\resizebox{\hsize}{!}{\includegraphics[bb= 30 22 527 463]{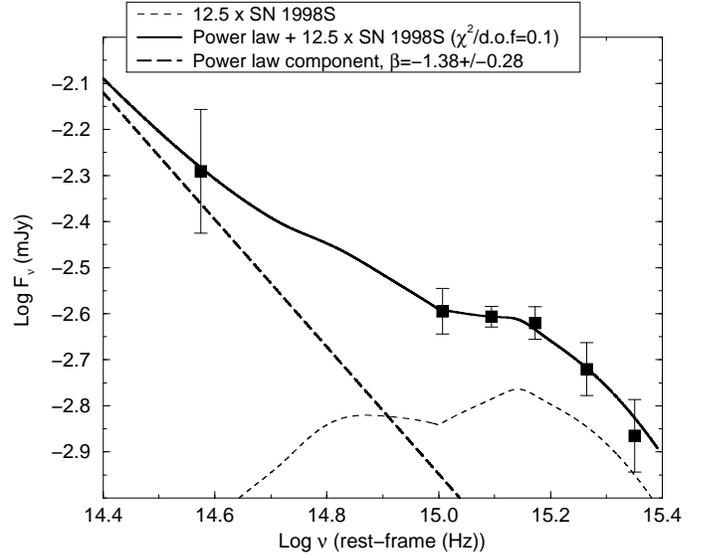}}
\caption{\label{sed2} The high redshift SN scenario.  As in Fig.~\ref{sed1}
the filled squares display the $UBVRIK^{\prime}$-band SED around March 17.0
UT.  The solid curve displays the satisfactory ($\chi^2/{\rm d.o.f} = 0.1$)
result obtained with the SN~1998S  template when a $F_{\nu} = const. \times
\nu^{\beta} +  k \times  F_{\nu}^{SN}$ function is  fitted if  $z=1.73$ and
$k=12.5$.   The long dashed  straight line  shows the  underlying power-law
spectrum  ($\beta=-1.38\pm0.28$)  whereas the  dashed  curve  shows the  SN
component.}
\end{center}
\end{figure}

\subsection{Possible SN interpretations}

\subsubsection{The SED}
\label{SEDdiscussion}

The   SED  around   17.0  March   UT   shows  a   blue  optical   afterglow
($U-B=-0.56\pm0.18$,          $B-V=0.31\pm0.13$,         $V-R=0.09\pm0.09$,
$R-I=0.29\pm0.12$, corrected for Galactic extinction), in comparison to the
mean  colours   measured  for   other  optical  afterglows   ($\langle  B-V
\rangle=0.47\pm0.17$,  $\langle   V-R  \rangle=0.40\pm0.13$,  $\langle  R-I
\rangle=0.46\pm0.18$; $\check{\rm S}$imon et al. \cite{Simo01}).

In order to  fit our optical/NIR SED, we  considered two spectroscopic
templates; SN~1998bw (moderately blue, Patat et al. \cite{Pata01}) and
SN~1998S  (very  blue,  Fassia  et al.   \cite{Fass01}).   The  fitted
expression  is given  by; $F_{\nu}  = const.   \times \nu^{\beta}  + k
\times F_{\nu}^{SN}$, where $F_ {\nu}^{SN}$  is the SN template at the
GRB redshift  and $k$  an amplitude.  As  shown in  Fig.~\ref{sed1}, a
dimmed SN~1998bw  template at $z\sim0.1$  improves the quality  of the
fit with respect  to the pure power-law ($\chi^{2}/{\rm  d.o.f} = 0.3$
vs.   $\chi^{2}/{\rm  d.o.f}  =   0.8$),  yielding  a  spectral  index
$\beta=-0.59  \pm   0.21$.   However,  at   $z\sim0.1$  the  SN~1998bw
amplitude  has to  be dimmed  by  a factor  of $\sim50$  to reach  the
observed  optical  flux  level,  implying  a  $B$-band  absolute  peak
magnitude  of: $M_{B}\sim-15$.   This would  imply a  very subluminous
type Ic  SN (e.g., Richardson et al.   \cite{Rich02}).  The moderately
blue SN~1998bw template  is unable to reproduce the  observed blue SED
beyond $z\sim0.2$.   We note that the SN~1998bw  spectrum was extended
in the $\lambda < 3600$\AA~  rest-frame UV domain adopting a power-law
$F^{SN}_{\nu} \sim \nu^{-2.8}$ spectrum as previously assumed by other
authors (Bloom et al. \cite{Bloo99}; Lazzati et al. \cite{Lazz01}).

The  UV  rest-frame  flux  of  SN~1998S has  been  obtained  combining
HST(+STIS)  observational spectra  (Lentz et  al.   \cite{Lent01}) and
black  body   extensions  to  the  optical  spectra   (Fassia  et  al.
\cite{Fass01}).   The  black  body  temperatures  have  been  obtained
interpolating in time  the SN~1998S photospheric temperatures (Anupama
et al.  \cite{Anup02}; Fassia et al. \cite{Fass00}).

With the  bluer SN~1998S template it is  possible to fit the  SED at larger
redshifts. The best  fit based on SN~1998S is  reached at $z\sim1.7$, where
the  fit  quality  is  considerably  improved in  comparison  to  the  pure
power-law fit ($\chi^{2}/{\rm  d.o.f}=0.1$ vs.  $\chi^{2}/{\rm d.o.f}=0.8$,
see  Fig.~\ref{sed2}).   In  this   case  the  derived  spectral  index  is
$\beta=-1.38\pm 0.28$.  The  amplitude of the best fit  is $\sim12.5$ times
SN~1998S, which  implies $M_{B} \sim -21.5$ (see  Fig.~\ref{sed2}).  To our
knowledge only the exceptionally bright  Ic-type SN 1999as has reached such
high luminosity ($M_{B} \sim -21.4$; Hatano et al. \cite{Hata01}).

Therefore, both  a low  redshift and a  high redshift SN  scenario are
able to fit  the blue and bumpy optical/NIR SED  around March 17.0 UT.
Note that we do not exclude  SN solutions based on SNe at intermediate
redshifts (i.e., $0.2<z<1.7$).

Two additional  complementary arguments support a  substantial SN flux
contribution around 11.5 days after the burst: {\it i)} The colours of
the afterglow  roughly agree with  the ones measured for  several blue
SNe  close  to  their  lightcurve  peaks: SN~1993J  (Richmond  et  al.
\cite{Rich94}), SN~1998S (Fassia et  al.  \cite{Fass01}), {\it ii)} at
low  redshift the  addition  of a  black  body term  to the  power-law
spectrum improves the  fit ($\chi^2/{\rm d.o.f}=0.4$ vs.  $\chi^2/{\rm
d.o.f}=0.8$, see Fig.~\ref{sed1}), yielding rest-frame temperatures of
$T\sim9200$~K,  compatible  with  the high  photospheric  temperatures
measured in SNe.  This scenario  resembles the SED of the low redshift
GRB~011121 at  $t-t_{0}\sim13.5$ days (being $t_{0}$ the  epoch of the
gamma-ray event),  which was reproducible  with a black  body template
having a temperature of $T\sim6300$~K (Greiner et al. \cite{Grei03}).

\subsubsection{The lightcurve}
 \label{lightcurvediscussion}

We have based  our lightcurve study on three  type Ic templates (SN~1998bw:
Galama et al.  \cite{Gala98}, McKenzie \& Schaefer \cite{McKe99}; SN~1994I:
Lee  et al.   \cite{Lee95},  Richmond et  al.   \cite{Rich96}, Tsvetkov  \&
Pavlyuk \cite{Tsve95};  SN~2002ap: Foley  et al.  \cite{Fole03},  Yoshii et
al.   \cite{Yosh03},  Gal-Yam  et  al.  \cite{GalY02},  Mattila  \&  Meikle
\cite{Mati02}, Hasubick \cite{Hasu02}) and one type II SN (SN~1998S: Fassia
et al.   \cite{Fass00}, Li et  al.  \cite{Li02}, Liu et  al.  \cite{Liu00},
Schaefer \& Roscherr \cite{Scha98}).

\begin{figure}[t]
\begin{center}
\resizebox{\hsize}{!}{\includegraphics[bb= 62 39 527 455]{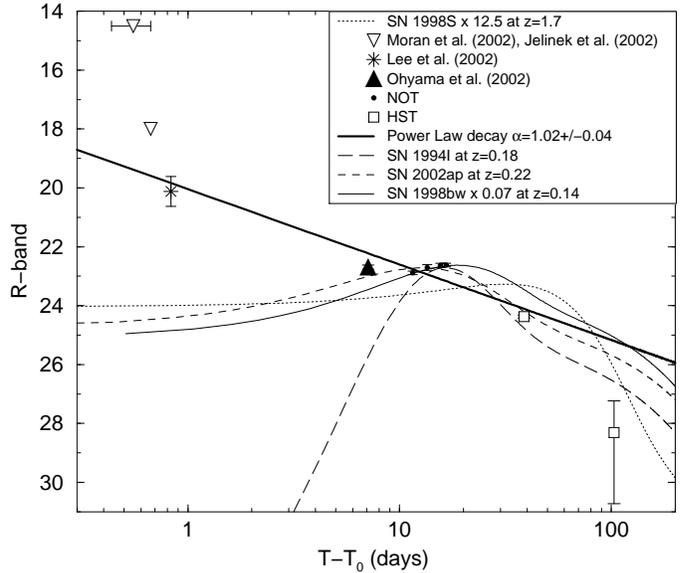}}
\caption{\label{lightcurve1} $R$-band lightcurve of the GRB 020305 OA, once
 the contribution of the host  galaxy has been subtracted. The small filled
 circles show  the NOT  data points.  The  empty squares give  the $R$-band
 magnitudes inferred from the HST  observations.  The other two data points
 were  taken  from Lee  et  al.  (\cite{Lee02};  star)  and  Ohyama et  al.
 (\cite{Ohya02};  filled  triangle).   The   upper  limits  by  Jelinek  et
 al. (\cite{Jeli02})  and Moran et  al. (\cite{Mora02}) are  represented by
 the two  empty triangles.   The thick solid  straight line shows  the best
 power-law    fit   obtained   when    $\alpha=1.02\pm0.04$   ($\chi^2/{\rm
 d.o.f}=34.9$).  As it  is shown the lightcurve shows  a clear bump between
 $\sim11.5$ and  $\sim16.7$ days after  the GRB, clearly deviated  from the
 pure power-law fit.   The plot shows three SN  lightcurve templates at low
 redshift ($z\sim  0.2$): SN~1998bw  (solid), SN~2002ap (dashed  curve) and
 SN~1994I  (long dashed).  It  also shows  a SN  template at  high redshift
 ($z\sim1.7$): SN~1998S (dotted).  The lower redshift templates give better
 fits than the  high redshift one, however the late  time fast decay (given
 by the two HST points of the plot) is difficult to match at any redshift.}
\end{center}
\end{figure}

In Fig.~\ref{lightcurve1} we have  overplotted the lightcurves of these SNe
on our $R$-band  data points.  The SN~1998bw and  SN~1998S lightcurves have
been  dimmed  by a  similar  factor  to the  one  needed  to  fit the  SEDs
(Figs.~\ref{sed1} and \ref{sed2}, respectively).  At low redshift SN~1998bw
and SN~2002ap  fit the  peak reasonably well,  but the fast  decay SN~1994I
provides the  best match being the only  one able to explain  the first HST
data point.  We note that  SN~1994I-like lightcurves also provide good fits
for the lightcurves  of GRB~021211 (Della Valle et  al.  \cite{Dell03}) and
XRF~030723 (Fynbo et al.  \cite{Fynb04}).

The   SN~1998S   lightcurve   redshifted   to  $z=1.7$   peaks   too   late
($t-t_{0}\sim30$  days)  and  is  therefore highly  inconsistent  with  the
$R$-band  data  points.   Lower  redshifts  provide  better  fits  for  all
templates, but  the late time fast decay  (given by the two  HST points) is
difficult to  fit at any  redshift.  Even with  a SN~1994I template  at low
redshift the second HST data point  is not well fitted. However, a break in
the afterglow lightcurve would yield a lower total (SN + afterglow) flux at
the second HST  epoch, hence leading to a better  fit.  From the lightcurve
peak epoch we conclude that a high redshift SN scenario seems unlikely.

\subsection{A SN-less interpretation}

An alternative interpretation of  the $RI$-band SED decrement is based
on the  redshifted 2175~\AA~ broad absorption bump.  In this framework
the SED  shape is not  the result  of a SN,  but rather the  result of
features in the extinction law. The 2175~\AA~ bump is prominent in the
Milky  Way (MW),  moderate in  the  Large Magellanic  Could (LMC)  and
almost undetectable in the Small  Magellanic Cloud (SMC).  In order to
test this  possibility, the flux  densities corresponding to  our data
points  were fitted  with an  expression  in the  form; $F_{\nu}  \sim
\nu^{\beta}  \times   10^{-0.4  A_{\nu}}$,  where   $A_{\nu}$  is  the
extinction  in  magnitudes  at  rest-frame frequency  $\nu$.   We  have
considered  the three  extinction  laws given  by Pei  (\cite{Pei92}),
i.e., for  the MW,  LMC and SMC.   $A_{\nu}$ has been  parametrised in
terms  of   $A_{V}$,  so  the  fit  determines   $A_{V}$  and  $\beta$
simultaneously.  Given  that the OA distance is  unknown, the redshift
has been ranged from $z=0$ to $z=2.8$.

\begin{figure}[t]
\begin{center}
\resizebox{\hsize}{!}{\includegraphics[bb= 30 20 527 471]{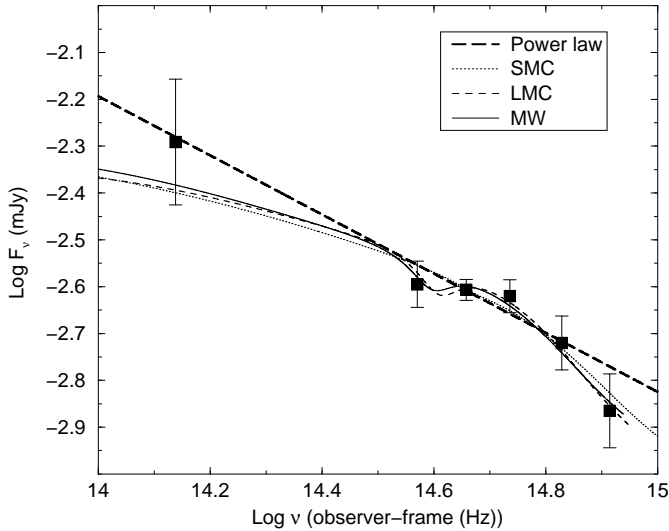}}
\caption{\label{sednoSN}  The  $UBVRIK^{\prime}$-band  photometry  measured
  around March  17.0 UT and  the best-fitting SED solutions  (not including
  any SN  component).  The three thin  curves show the  results of assuming
  different  extinction laws  (MW, LMC  and  SMC).  The  thick long  dashed
  straight  line  displays a  pure  power-law.   The  best fitted  solution
  ($\chi^2/{\rm  d.o.f} = 0.1$)  corresponds to  a MW  extinction law  at a
  redshift $z = 2.5$ (solid curve).  The absorption between the $R$ and the
  $I$-band  might  be  due  to  the  redshifted  graphite  2175~\AA~  broad
  signature,  prominent in  the MW,  faint in  the LMC  (dashed  curve) and
  almost absent in the SMC extinction law (dotted curve).}
\end{center}
\end{figure}

Table~\ref{NoSNSED1} shows that  the best fit solution ($\chi^{2}/{\rm
d.o.f}=0.1$)  is  reached  with   a  MW  extinction  law  at  $z=2.5$,
consistent with a power-law  spectrum with $\beta=-0.35\pm0.23$ and an
extinction $A_{\rm  V}=0.25\pm0.14$ (see Fig.~\ref{sednoSN}).   The MW
is the only extinction law providing a spectral index fully consistent
(within  error bars)  with $\beta<0$,  and therefore  agrees  with the
$\beta$ values  measured in OAs to  date.  For the MW  any redshift in
the range $2.2<z< 2.8$ provides values of $\chi^{2}/{\rm d.o.f} < 1$.

However,  this  interpretation ``per  se''  can  not  account for  the
simultaneous  rebrightening of  the lightcurve  12--16 days  after the
burst,  being necessary an  additional mechanism  to explain  it.  One
possible  solution would  be to  describe the  lightcurve bump  in the
context  of the  two-component  jet model,  invoked  to interpret  the
rebrightenings measured in  GRB\,030329 (Berger et al.  \cite{Berg03})
and XRF~030723 (Huang et al. \cite{Huan04}).

This scenario is  opposite to the OA SEDs measured  to date, where for
all cases the metal-rich MW  extinction law gives poor fits (Jensen et
al.   \cite{Jens01};  Fynbo  et  al.  \cite{Fynb01};  Holland  et  al.
\cite{Holl03};  Jakobsson et al.   \cite{Jako04}).  Thus,  we consider
this interpretation less likely than the SN interpretation, however it
can not be excluded.

\begin{table}
\begin{center}
\caption{Parameters derived  from the SED fits  when a SN  component is not
considered.  The  table shows  the fit results  for three  extinction laws,
given  in  the first  column.   The rest  of  columns  display the  derived
redshift  ($z$),  extinction   ($A_{V}$),  spectral  index  ($\beta$),  and
goodness of the solution, given by $\chi^{2}/{\rm d.o.f}$.}

\begin{tabular}{lcccc}
\hline
Ext. Law& $z$ &   $A_{V}$  & $\beta$    &$\chi^{2}/{\rm d.o.f}$ \\
\hline      
 SMC & $1.9$ & $0.16\pm0.15$ &$-0.13\pm0.51$ & $0.8$ \\
 LMC & $2.4$ & $0.32\pm0.21$ &$ 0.10\pm0.50$ & $0.3$ \\ 
 MW  & $2.5$ & $0.25\pm0.14$ &$-0.35\pm0.23$ & $0.1$ \\
\hline
\label{NoSNSED1}
\end{tabular}
\end{center}
\end{table}

\section{Conclusions}
\label{conclusions}

 In this paper optical ground-based and HST imaging have been reported
for the  GRB 020305 afterglow.   The HST imaging revealed  an extended
host  galaxy  with  $R=25.17\pm0.14$,   which  might  be  part  of  an
interacting system.

 The lightcurve of  the afterglow, constructed with our  data and with data
of other authors, shows a rebrightening $12-16$ days after the GRB.  On the
other hand,  the $UBVRIK^{\prime}$-band SED constructed around  the time of
the expected  SN peak shows a blue  and bumpy SED which  might deviate from
the  power-law  fit  with  $\beta=-0.63\pm0.16$.   The  afterglow  $U$-band
detection imposes an  upper limit to the GRB redshift  of $z \lesssim 2.8$,
inconsistent with previous redshift estimates (Atteia \cite{Atte03}).

 We   have  considered  several   interpretations  that   may  account
simultaneously  for the  lightcurve rebrightening  and the  bumpy blue
spectral energy distribution.  The most natural scenario is based on a
SN associated to the GRB.

 Our $UBVRIK^{\prime}$-band SED can not precisely constrain the redshift of
the GRB. Both a low ($z  \lesssim 0.2$) and a high redshift ($z\sim1.7$) SN
solutions can  match the SED  reasonably well.  The high  redshift scenario
has  the (only)  advantage of  being able  to explain  easily the  low host
galaxy luminosity, which seems  unlikely for $z\sim0.1$.  However, the high
redshift  scenario  requires  an   extraordinarily  bright  SN  ($M_B  \sim
-21.5$). Contrary to  this, the absolute peak magnitude  derived in the low
redshift SN solution is more  realistic, especially given that the fraction
of SNe that  are subluminous ($M_B > -15$) appears to  be higher than $1/5$
(Richardson  et al.  \cite  {Rich02}).  Furthermore,  the high  redshift SN
scenario is not able to  fit the $R$-band lightcurve, and especially around
the rebrightening.  In conclusion, we consider the low redshift SN scenario
the most likely interpretation of the data.

 If the  GRB~020305 SED/lightcurve  bumps have a  SN origin,  then our
 data would  support the diversity of  the SNe related  to GRBs, since
 the extensively  used SN~1998bw  template does not  necessarily yield
 the best fits.

 In an alternative SN-less interpretation  the bumpy SED is the result
 of structure  present in the extinction  curve. A fit  with a MW-like
 extinction law leads to an estimated redshift of $z\approx2.5$.  This
 interpretation can not naturally describe the simultaneous lightcurve
 rebrightening, requiring an extra explanation for the lightcurve bump
 (given for  instance by  the two-component jet  model), and  hence we
 consider it less likely than the SN interpretations.

 The  determination of  the host  galaxy redshift,  either via  a deep
 spectrum or based on  a photometric redshift would drastically reduce
 the many free parameters of the present study.

\begin{acknowledgements}
 
Some of  the data  presented here  have been taken  using ALFOSC,  which is
owned  by the  Instituto de  Astrof\'{\i}sica de  Andaluc\'{\i}a  (IAA) and
operated at  the Nordic Optical  Telescope under agreement between  IAA and
the  Niels Bohr  Institute. We  are  grateful to  the staff  at the  Nordic
Optical Telescope  for excellent  support.  This work  is supported  by the
Danish  Natural  Science  Research  Council  (SNF).     We  acknowledge
D.L.  Tucker, B.C.  Lee, and  G.  Kosugi  for valuable  information  on the
seeing  conditions  and  the  calibration  stars  used  by  Ohyama  et  al.
(\cite{Ohya02}).   Support  for Proposal  GO  9074  was  provided by  NASA
through  a grant  from  the  Space Telescope  Science  Institute, which  is
operated  by the  Association of  Universities for  Research  in Astronomy,
incorporated under  NASA contract  NAS5-26555.  This research  is partially
supported  by  the  Spanish  Ministry  of  Science  and  Education  through
programmes ESP2002-04124-C03-01 and  AYA2004-01515 (including FEDER funds).
We thank our anonymous referee for useful and constructive comments.

\end{acknowledgements}

\end{document}